\documentclass[conference]{IEEEtran}
\IEEEoverridecommandlockouts

\usepackage{amsmath,amssymb,amsfonts}
\usepackage{cite}
\usepackage{dsfont}
\usepackage{graphicx}
\usepackage{numprint}
\usepackage{upgreek}

\usepackage{coverpage}

\npdecimalsign{\ensuremath{.}}

\def\BibTeX{{\rm B\kern-.05em{\sc i\kern-.025em b}\kern-.08em
    T\kern-.1667em\lower.7ex\hbox{E}\kern-.125emX}}
    
\tdtitle{Modeling the Spatial Distributions of Macro Base Stations with Homogeneous Density: Theory and Application to Real Networks}
\tdauthor{Q. Gontier$^{(1*)}$, C. Tsigros$^{(2)}$, F. Horlin$^{(1)}$, J. Wiart$^{(3)}$, C. Oestges$^{(4)}$, Ph. De Doncker$^{(1)}$}
\source{$^{(1)}$Wireless Communications Group, Universit\'e Libre de Bruxelles, Brussels, Belgium\\
$^{(2)}$Département Technologies et Rayonnement, Brussels Environnement, Brussels, Belgium\\
$^{(3)}$LTI, Telecom Paris, Chaire C2M, Institut Polytechnique de Paris, Palaiseau, France\\
$^{(4)}$ICTEAM Institute, Universit\'e Catholique de Louvain, Louvain-la-Neuve, Belgium\\} 
\tdnumber{74}
\location{Bologna, Italy}
\dates{8-11 February}
\address{Avenue Franklin Roosevelt 50 - CP 165/81, 1050 Ixelles, Belgium}
\phone{}
\fax{}
\email{quentin.gontier@ulb.be}

\begin{document}

\title{Modeling the Spatial Distributions of Macro Base Stations with Homogeneous Density: Theory and Application to Real Networks\\
\thanks{This work was supported in part by Innoviris under the STOEMP-EMF grant.}
}


\makecoverpage
\cleardoublepage

\maketitle
\begin{abstract}
Stochastic geometry is a highly studied field in telecommunications as in many other scientific fields. In the last ten years in particular, theoretical knowledge has evolved a lot, whether for the calculation of metrics to characterize interference, coverage, energy or spectral efficiency, or exposure to electromagnetic fields. Many spatial point process models have been developed but are often left aside because of their unfamiliarity, their lack of tractability in favor of the Poisson point process or the regular lattice, easier to use. This article is intended to be a short guide presenting a complete and simple methodology to follow to infer a real stationary macro antenna network using tractable spatial models. The focus is mainly on repulsive point processes and in particular on determinantal point processes which are among the most tractable repulsive point processes. This methodology is applied on Belgian and French cell towers. The results show that for all stationary distributions in France and Belgium, the best inference model is the $\beta$-Ginibre point process.\\
\end{abstract}


\section{Introduction}

In recent years, cellular networks have been modeled stochastically to take advantage of the benefits of stochastic geometry (SG). Indeed, there is a great lack of accuracy when a network is modeled as a regular network. The other option would be to know the exact location of every antenna in the network, but then calculating the metrics mathematically without involving simulations becomes impossible. SG enables the representation of large networks and the calculation of statistics and metrics on exposure \cite{GontierAccess}, SINR \cite{baccellimetrics}, energy correlation \cite{ECC}...

SG and its inherent spatial point processes (PPs) are widely used in a lot of fields (biology \cite{cressie1993, Cauchy_Ghorbani, Othmer88}, ecology \cite{thompson55}, geology \cite{volcanism}, seismology \cite{gardner74}, astronomy \cite{BABU1996311}, etc.). Although a whole theory has been developed, there is a kind of reluctance to use advance PPs to model experimental patterns, especially for cellular networks. Authors usually settle for the regular hexagonal lattice \cite{hexagonal2016}, the perfect square lattice \cite{InterferenceLargeNetworkHaenggi} or the Poisson point process (PPP) \cite{Lee2013, GontierAccess} which leads to very tractable analytical expressions but does not faithfully model the geometry of real networks. We will show in this paper that it is possible to better model cellular network while maintaining the tractability of the expressions. However, several works succeeded in fitting real antenna patterns using different methods and models. According to \cite{BGPP_Gomez}, Paris' network could be modeled as a $\beta$-Ginibre PP ($\beta$-GPP) in 2014. Other European cities were modeled using a log-Gaussian Cox process (LGCP) \cite{Kibilda2016}, a Strauss process (SP) \cite{Anjin2013}, a Geyer saturation process \cite{RIIHIJARVI2012858} a Cauchy PP (CPP) or hybrid models \cite{Zhang_2021}... In the United States, Li et al. \cite{li2014statistical, Baccelli_DPP2} used determinantal PPs (DPPs) as an alternative to Gibbs PPs (GPPs) to capture the repulsive geometry of the Houston and Los Angeles base stations. 

These works, although remarkable for some, often model a certain type of environment (usually urban). In addition, the number of test-models used is generally quite small. Several models are used because user-friendly libraries exist in the \texttt{R} libraries. These test-models are not always tractable. When the PP is not tractable, some works approximate mathematical expressions, or use upper or lower bounds, or approximate a realization of the PP by a PPP or other more tractable PPs \cite{Deng2015HeterogeneousCN, IDT}.

This article aims to be a quick guide for PP inference in macro cellular networks. We propose to summarize the main tractable models and their characteristics. Here, tractable means that mathematical expressions can be found for at least the summary statistics and the moment-generating function. Definitions and empirical examples of summary statistics, which are mainly distance spacing and correlation functions, are also given in this paper. We give some hints on how to better account for edge effects.

We propose an approach to infer real datasets with spatial homogeneity in urban, suburban or rural areas. We remind the notions of stationarity, isotropy, complete spatial randomness and interaction between points.
\section{Spatial Point Processes}
\label{sec:spatialPP}
Characterization of a PP, as in classical statistics, consists in determining information from a finite subset of the PP, called sample, whose boundaries form a window. This can be done by means of sample statistics. Sample statistics correspond to any scalar value computed from the sample data. For example, the average or median distance from a random point to the nearest antenna or the nearest distance between two antennas is a measure of interest. In addition, summary functions and statistical tests are often sought to better analyze the distribution of points. Depending on the relationship between the points, a PP model can be fitted to the data, which is called spatial statistical inference. These models are characterized by different parameters to adjust the strength of the relationships between the points, the scale of variability, the density of parent and offspring (children) points in the case of clusters... 

The complete analysis of a PP requires computing the expectation of counts of points, pairs of points, triplets of points, etc., corresponding to different orders of moment quantities. For example, intensity is a first order moment quantity since it requires counting the number of points in a given region of space and is intrinsically related to an average. An example of a second-order moment property is the interaction between points, the stochastic dependence between two points, characterized by the $K$-function defined below.

Let us introduce some notations before going deeper into the theory. Let $x_i$ be the point of a spatial point process $\Psi$ and $u$ a random point which does not belong to $\Psi$ in general. $d(x_i, x_j) = d_{ij} = ||x_i-x_j||$ is the pairwise distance between the ordered pair of distinct points $x_i$ and $x_j$. The nearest-neighbor distance $d_i = d(x_i, \Psi \char`\\ x_i) = \min\limits_{j\neq i} d_{ij}$ is the distance from the data point $x_i$ to its nearest neighbor. $d(x_i, u)$ is the distance between $x_i$ and the random point $u$. Then, the empty-space distance $d(u, \Psi) = \min_j {||u-x_j|| \, |\,x_j \in \Psi}$ is the distance between a random point $u$ in the window to its nearest data point belonging to $\Psi$. 

\subsection{Invariance and Intensity Measure}

The intensity is the number of points per unit of dimensional measure. Since the considered PPs are bi-dimensional, the intensity corresponds to a 2D density of points. Starting with the analysis of the intensity already gives clues about spatial interactions. For example, a localized increase in intensity could be a hint to the presence of clusters.

A major point of interest is to know if the PP $\Psi = \left\{x\right\}$ is stationary, i.e. that the translated PP $\Psi_h = \left\{x + h\right\}$ has the same distribution as $\Psi$ for every $x \in \mathbb{R}^2$, because stationarity implies homogeneity of the density (the converse is in general not true). In the case of homogeneous intensity, the expected number of points falling in a given region is proportional to its area. Note that in this case, the empirical density of points $\widehat{\lambda}$ is an unbiased estimate of the true intensity $\lambda$. In the case of a homogeneous PPP (H-PPP), an estimate of the standard error of $\widehat{\lambda}$ is $\sqrt{\widehat{\lambda}/\nu_2(\mathcal{B})}$, where $\nu_2(\mathcal{B})$ is the area of the window $\mathcal{B}$.
 
In the case of a varying intensity $\lambda(u)$, one can try to fit it by a function. Various common models of the spatially-varying intensity exist and that can be used as a basis of inhomogeneous PPPs (I-PPPs), one of the only tractable inhomogeneous PPs \cite{R_spatstat}. Some specific models can be applied to an isotropic PP, a property of PPs $\Psi = \left\{x\right\}$ for which the rotated PP $\textbf{r}\Psi = \left\{\textbf{r} x\right\}$ has the same distribution as $\Psi$ for every rotation $\textbf{r}$ about the origin. Note that a stationary and isotropic PP is motion-invariant \cite{Daley1998}. The estimation of the inhomogeneous intensity can be performed non-parametrically using quadrat counting and kernel estimation \cite{R_spatstat}.

Note, however, that some PPs with significant aggregation behavior may unfairly lead one to believe that the node density is spatially inhomogeneous. To be sure, several stationarity tests can be run \cite{stationarity_tests13, stationarityKPSS}.

\subsection{The Poisson Point Process}

The PPP is one of the most widely used PPs, essentially due its convenient mathematical properties and its ability to realistically model PPs in many domains. As rigorously defined in \cite{tutorial}, a (spatial) H-PPP $\Phi \subset \mathbb{R}^d$ of uniform intensity $\lambda > 0 \left[\text{m}^{-d}\right]$ is a PP such that for every bounded close set $\mathcal{B}$ of dimensional measure $\nu_d(\mathcal{B})$, the number of points falling in $\mathcal{B}$, $n(\Phi \cap \mathcal{B})$, has a Poisson distribution with mean $\lambda\cdot\nu_d(\mathcal{B})$ and if $\mathcal{B}_{i \vert i = 1,...,m}$ are disjoint regions of $\mathbb{R}^d$, then the counts $n(\Phi \cap\mathcal{B}_i)_{ \vert i = 1,...,m}$ are independent. The Poisson distribution of $n(\Phi \cap\mathcal{B})$ is the consequence of the independence and uniform distribution of points of $\Phi$ (cf. \cite{kingman} for a rigorous demonstration). 

The points of a PPP do not demonstrate any attraction or repulsion relationship with each other. The central role of the PPP in SG studies comes from the fact that many models are built from it, and that many statistical tools are defined with respect to it.

\begin{figure*}[p]
    \centering
\begin{minipage}{.2\textwidth}
    \centering
    \includegraphics[width=\textwidth, trim={3cm, 9cm, 4cm, 10cm}, clip]{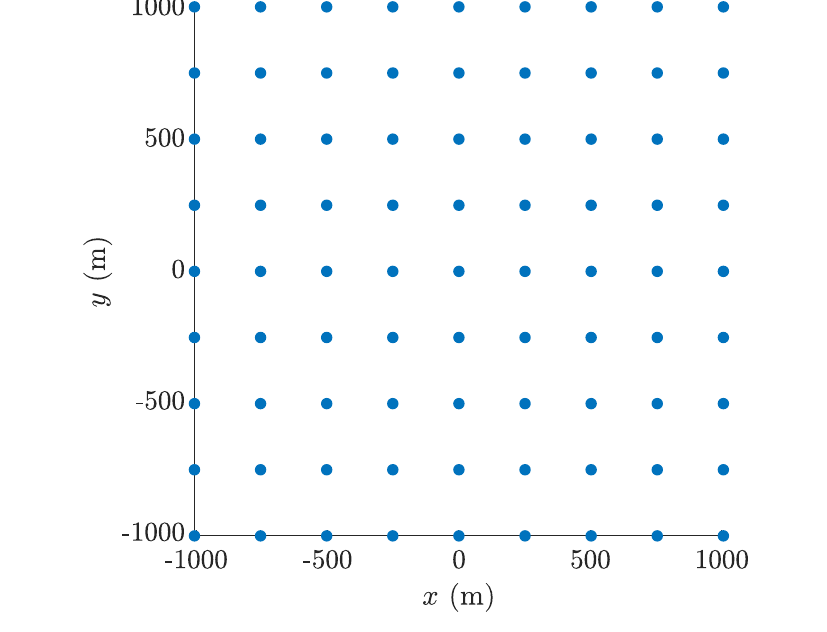}
    \label{fig:taxonomyregularNetwork.pdf}
\end{minipage}%
\begin{minipage}{.2\textwidth}
    \centering
    \includegraphics[width=\textwidth, trim={3cm, 9cm, 4cm, 10cm}, clip]{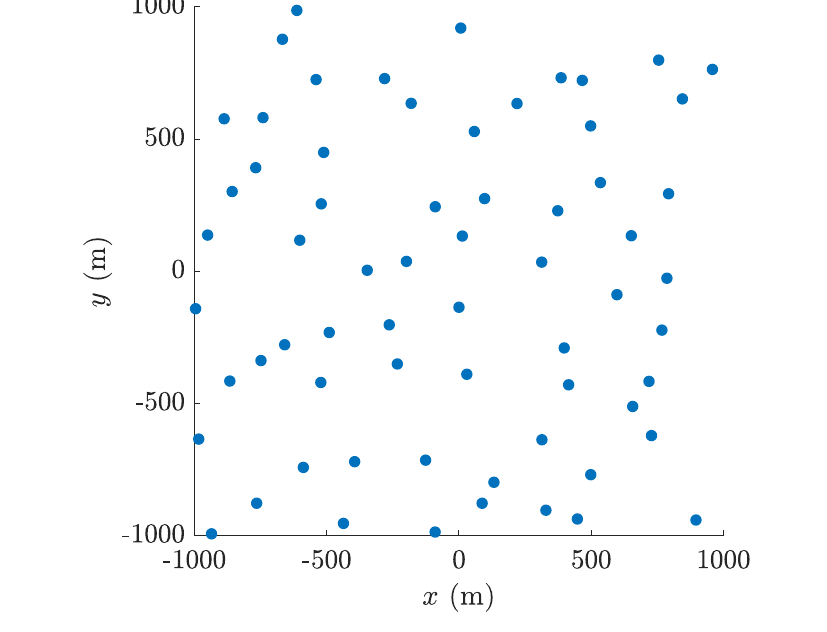}
    \label{fig:taxonomySmallRepulsion.pdf}
\end{minipage}%
\begin{minipage}{.2\textwidth}
    \centering
    \includegraphics[width=\textwidth, trim={3cm, 9cm, 4cm, 10cm}, clip]{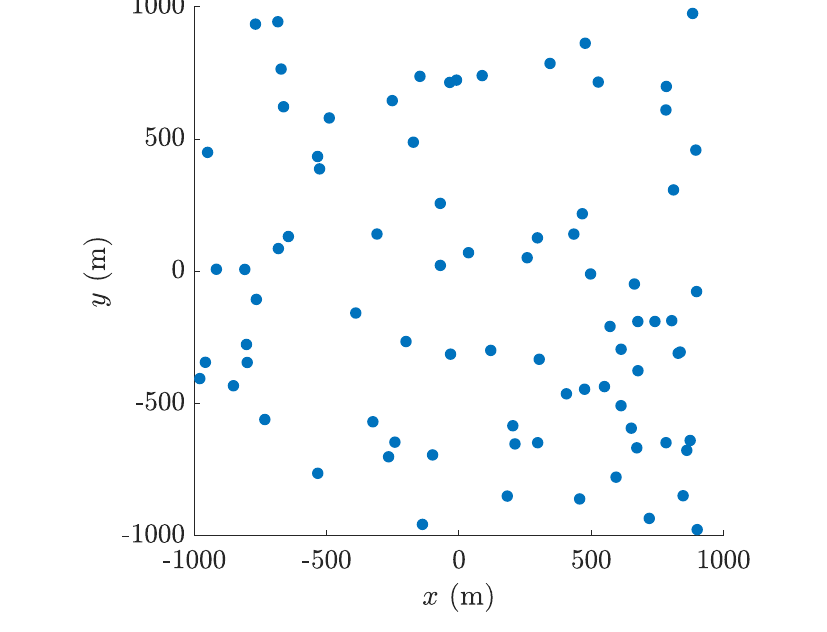}
    \label{fig:taxonomyPPP.pdf}
    \centering
\end{minipage}%
\begin{minipage}{.2\textwidth}
    \centering
    \includegraphics[width=\textwidth, trim={3cm, 9cm, 4cm, 10cm}, clip]{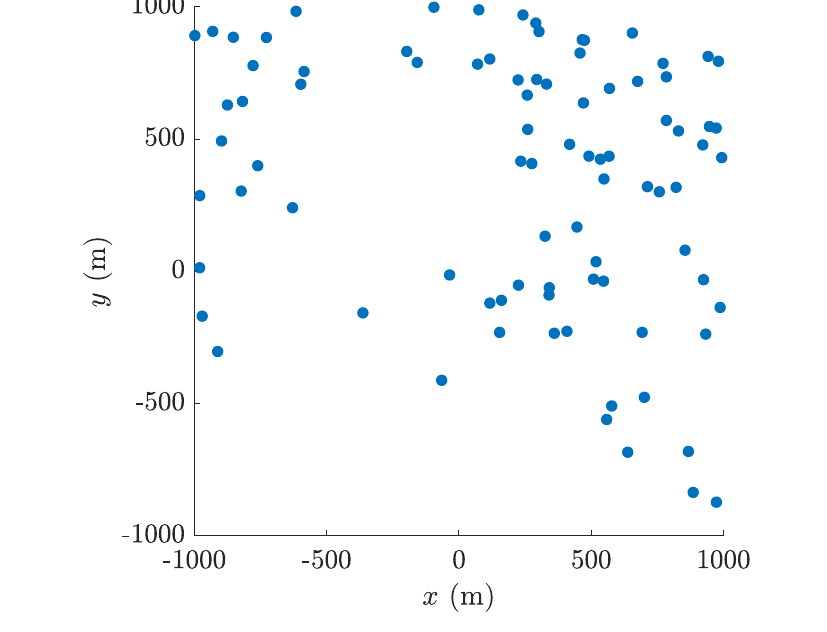}
    \label{fig:taxonomySmallAttraction}
\end{minipage}%
\begin{minipage}{.2\textwidth}
    \centering
    \includegraphics[width=\textwidth, trim={3cm, 9cm, 4cm, 10cm}, clip]{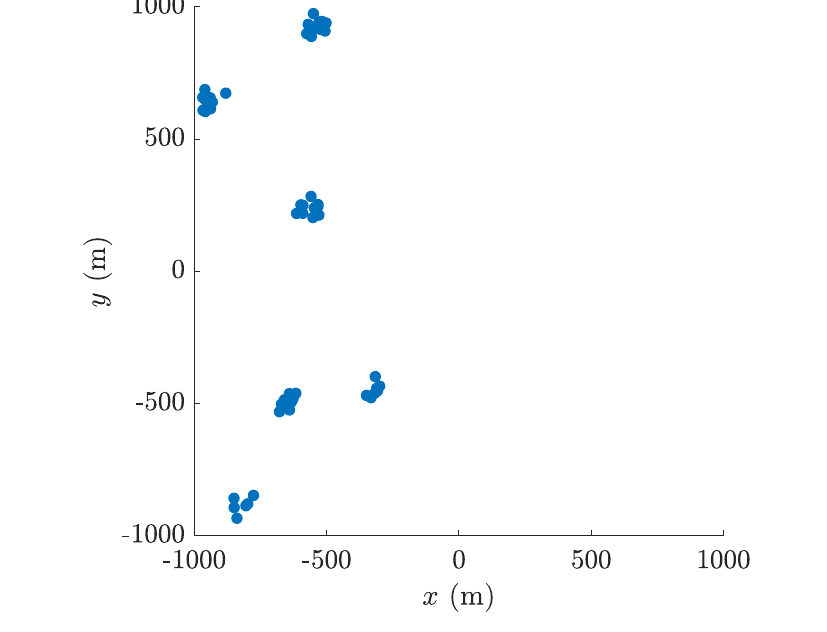}
    \label{fig:taxonomyHighAttraction}
\end{minipage}
\caption{\label{fig:PP_trichotomy}Illustration of PP trichotomy. Extreme left: Perfectly regular network. Center: H-PPP. Extreme right: Highly clustered pattern.}
\end{figure*}

\begin{figure*}[p]
    \centering
    \includegraphics[width=0.9\textwidth, trim={0cm, 23.0cm, 0cm, 4.6cm}, clip]{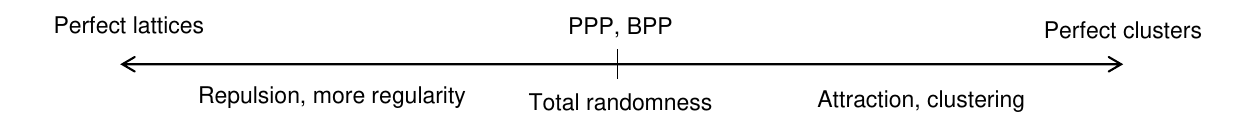}
    \label{fig:taxonomyarrows}
\end{figure*}

\begin{figure*}[p]
    \centering
\begin{minipage}{.2\textwidth}
    \centering
    \includegraphics[width=\textwidth, trim={3cm, 9cm, 4cm, 10cm}, clip]{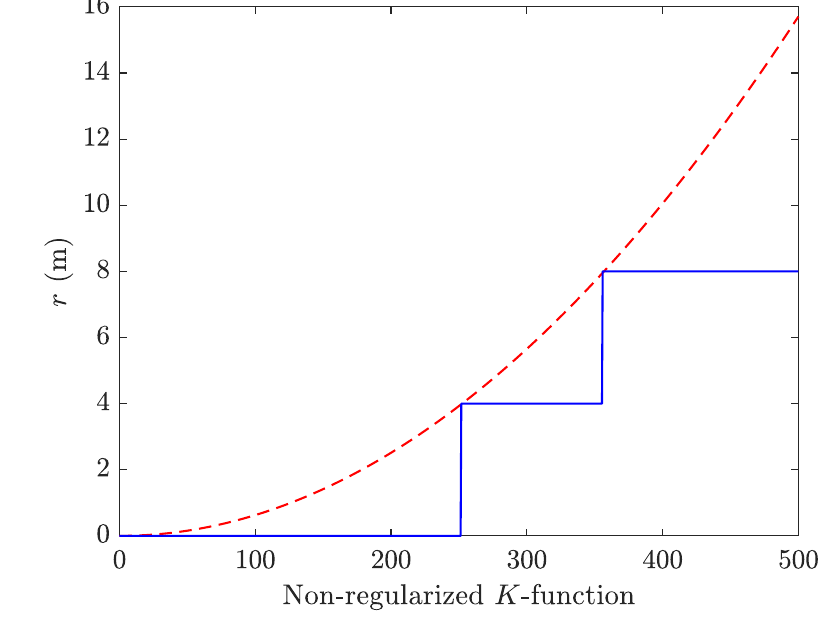}
    \label{fig:taxonomyregularNetwork_K.pdf}
\end{minipage}%
\begin{minipage}{.2\textwidth}
    \centering
    \includegraphics[width=\textwidth, trim={3cm, 9cm, 4cm, 10cm}, clip]{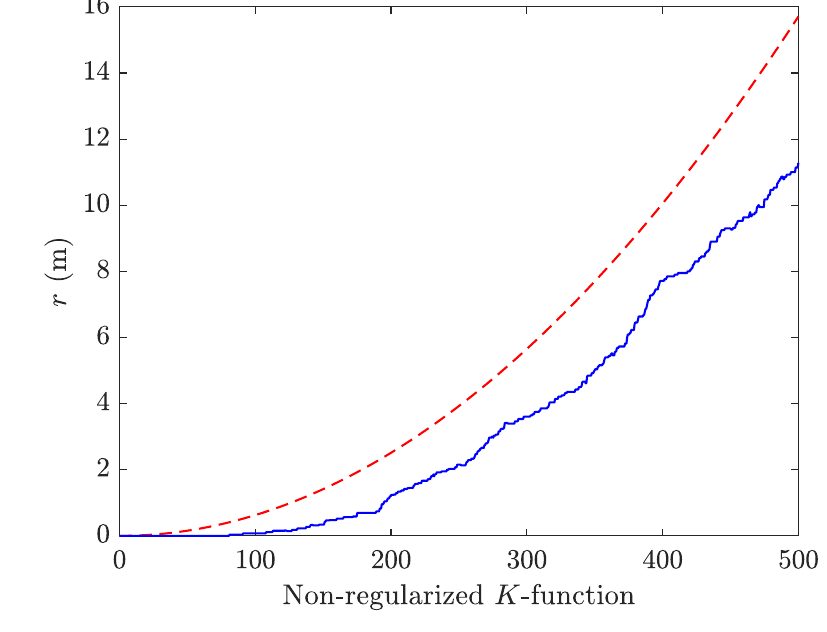}
    \label{fig:taxonomySmallRepulsion_K.pdf}
\end{minipage}%
\begin{minipage}{.2\textwidth}
    \centering
    \includegraphics[width=\textwidth, trim={3cm, 9cm, 4cm, 10cm}, clip]{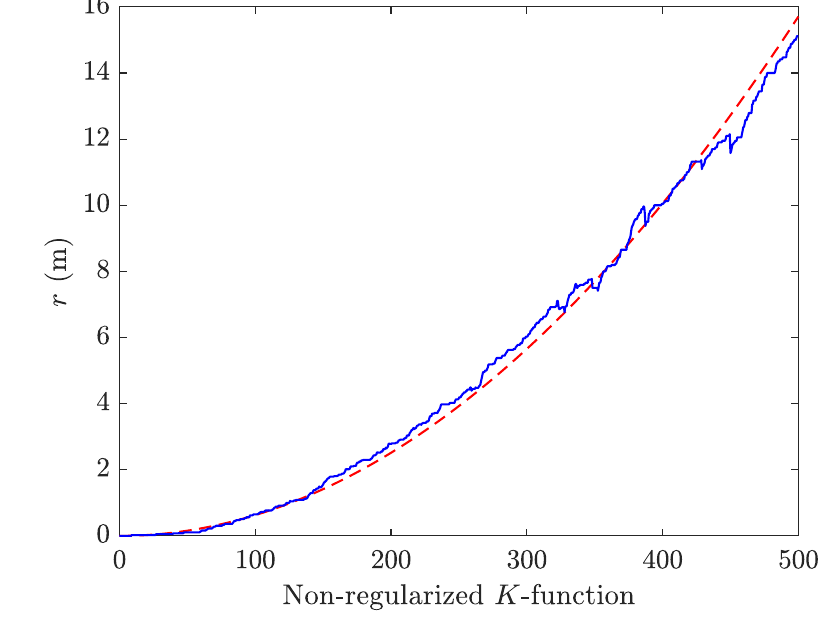}
    \label{fig:taxonomyPPP_K.pdf}
    \centering
\end{minipage}%
\begin{minipage}{.2\textwidth}
    \centering
    \includegraphics[width=\textwidth, trim={3cm, 9cm, 4cm, 10cm}, clip]{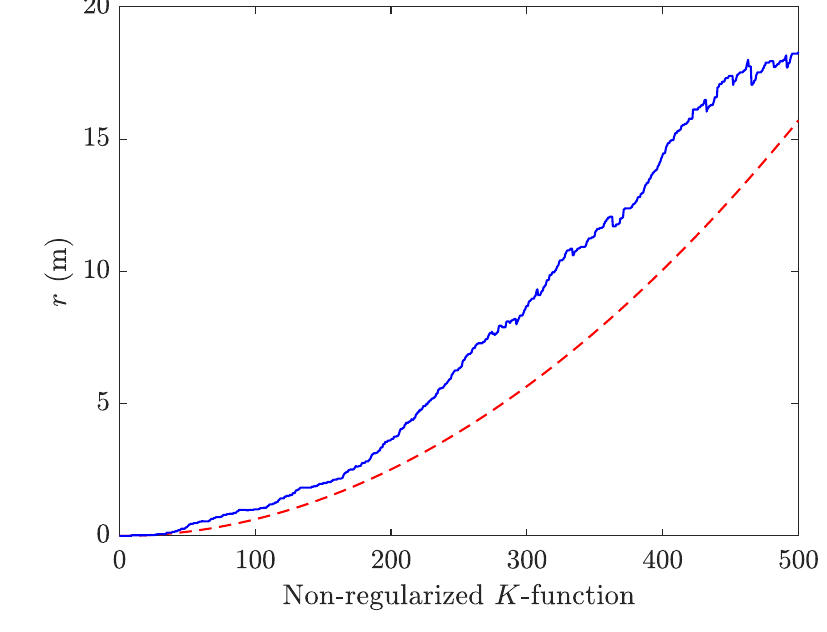}
    \label{fig:taxonomySmallAttraction_K}
\end{minipage}%
\begin{minipage}{.2\textwidth}
    \centering
    \includegraphics[width=\textwidth, trim={3cm, 9cm, 4cm, 10cm}, clip]{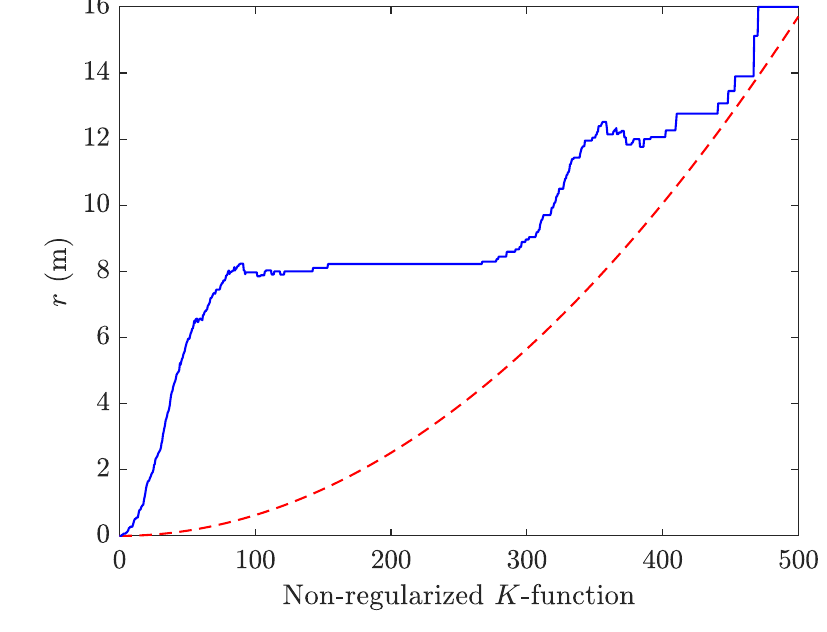}
    \label{fig:taxonomyHighAttraction_K}
\end{minipage}
\caption{\label{fig:PP_trichotomy_K}$K$-functions of the patterns of Figure \ref{fig:PP_trichotomy}. Extreme left: Perfectly regular network. Center: H-PPP. Extreme right: Highly clustered pattern.}
\end{figure*}

\begin{figure*}[p]
    \centering
\begin{minipage}{.2\textwidth}
    \centering
    \includegraphics[width=\textwidth, trim={3cm, 9cm, 4cm, 10cm}, clip]{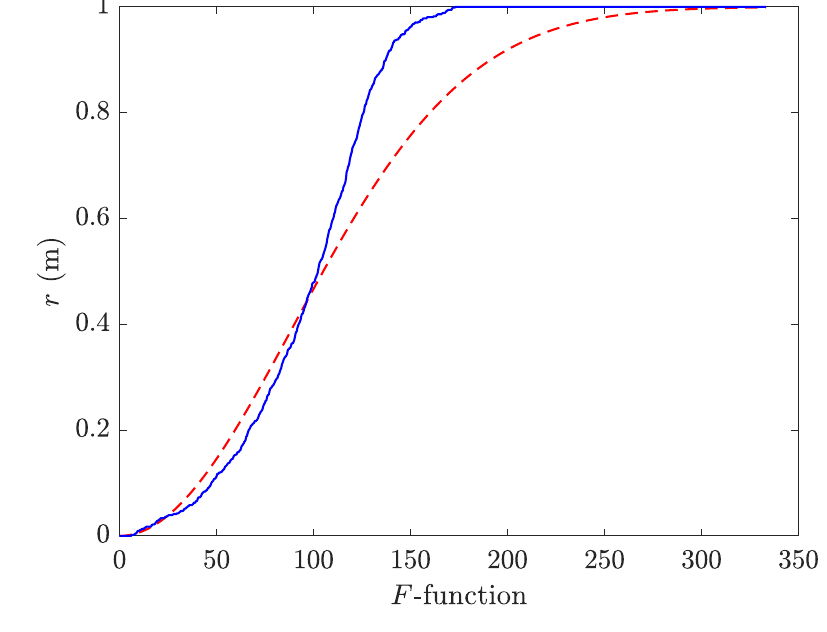}
    \label{fig:taxonomyregularNetwork_F.pdf}
\end{minipage}%
\begin{minipage}{.2\textwidth}
    \centering
    \includegraphics[width=\textwidth, trim={3cm, 9cm, 4cm, 10cm}, clip]{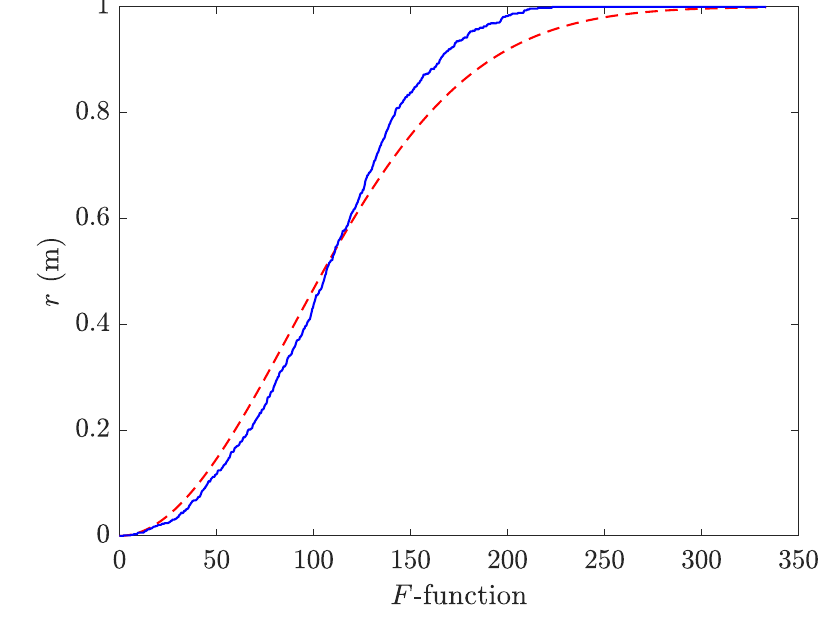}
    \label{fig:taxonomySmallRepulsion_F.pdf}
\end{minipage}%
\begin{minipage}{.2\textwidth}
    \centering
    \includegraphics[width=\textwidth, trim={3cm, 9cm, 4cm, 10cm}, clip]{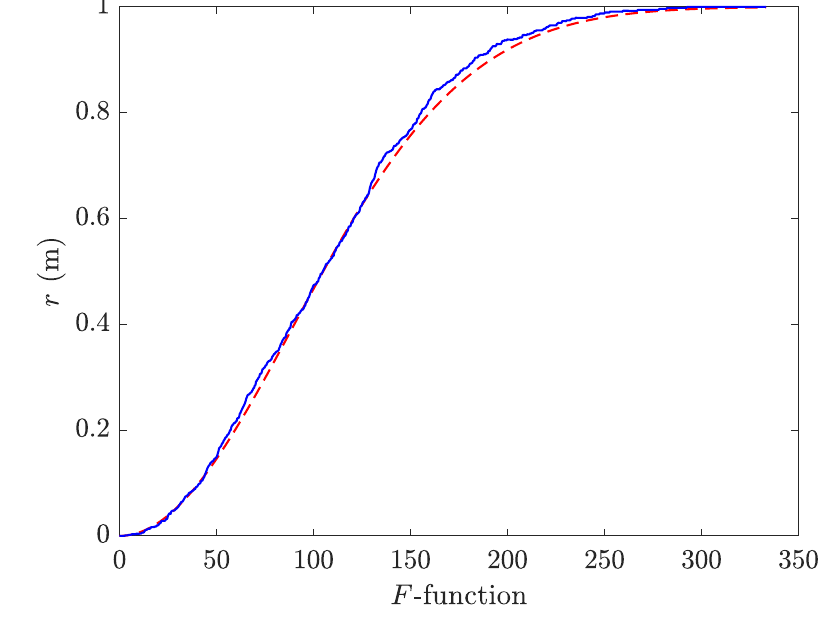}
    \label{fig:taxonomyPPP_F.pdf}
    \centering
\end{minipage}%
\begin{minipage}{.2\textwidth}
    \centering
    \includegraphics[width=\textwidth, trim={3cm, 9cm, 4cm, 10cm}, clip]{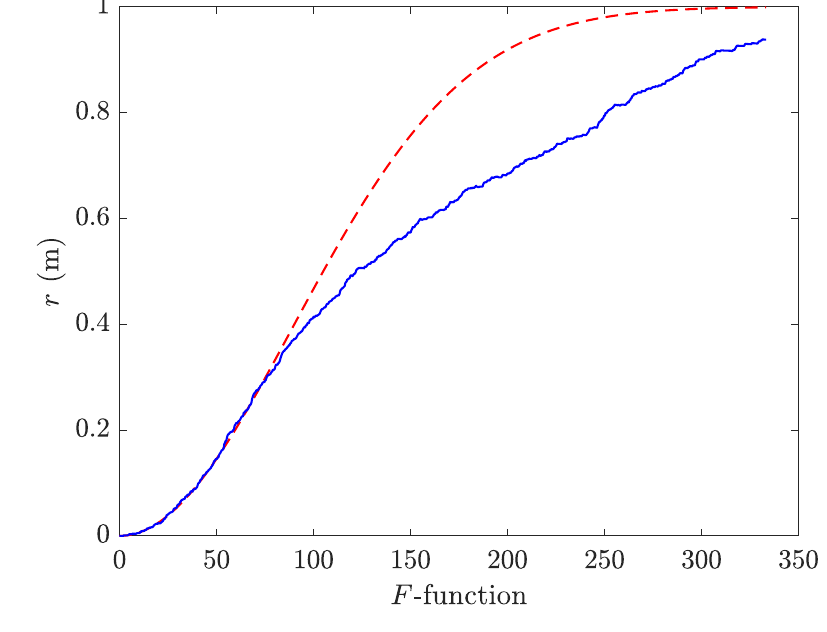}
    \label{fig:taxonomySmallAttraction_F}
\end{minipage}%
\begin{minipage}{.2\textwidth}
    \centering
    \includegraphics[width=\textwidth, trim={3cm, 9cm, 4cm, 10cm}, clip]{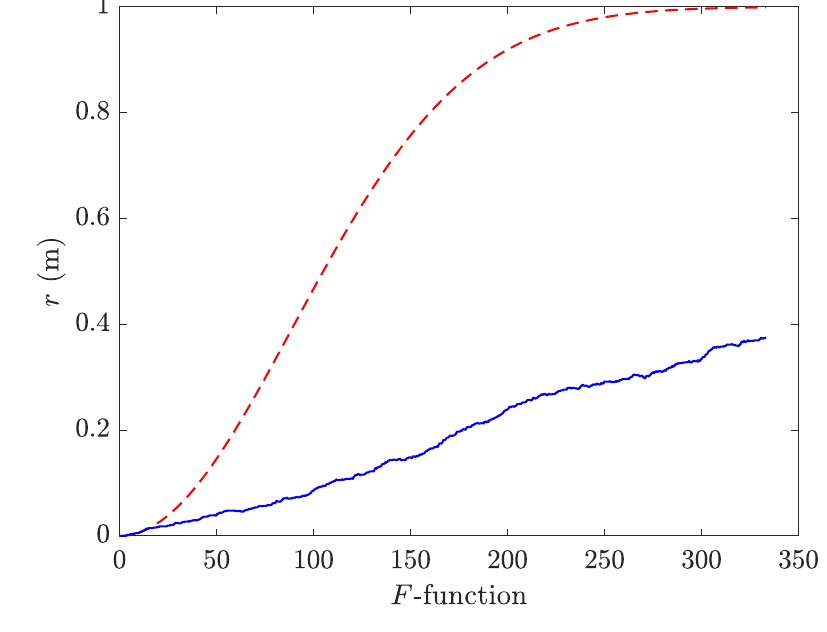}
    \label{fig:taxonomyHighAttraction_F}
\end{minipage}
\caption{\label{fig:PP_trichotomy_F}$F$-functions of the patterns of Figure \ref{fig:PP_trichotomy}. Extreme left: Perfectly regular network. Center: H-PPP. Extreme right: Highly clustered pattern.}
\end{figure*}

\begin{figure*}[p]
    \centering
\begin{minipage}{.2\textwidth}
    \centering
    \includegraphics[width=\textwidth, trim={3cm, 9cm, 4cm, 10cm}, clip]{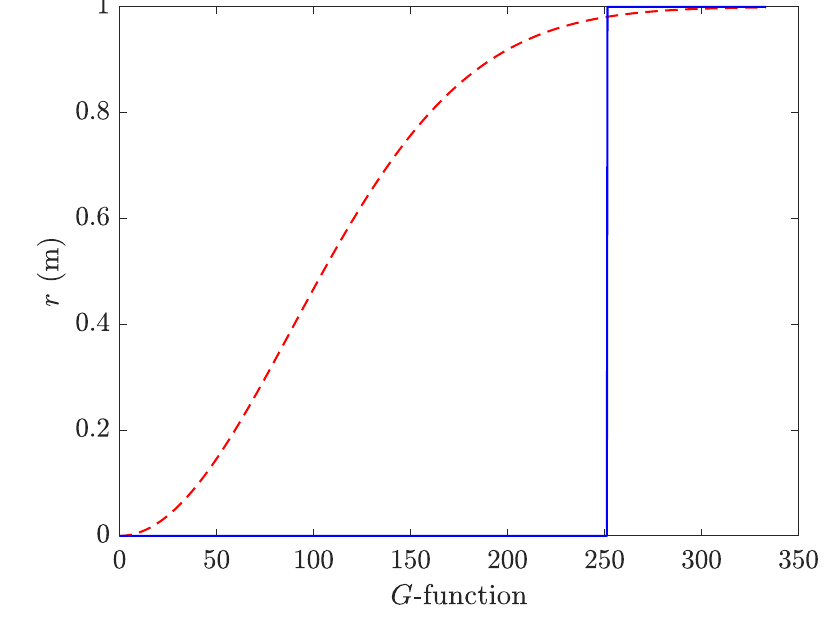}
    \label{fig:taxonomyregularNetwork_G.pdf}
\end{minipage}%
\begin{minipage}{.2\textwidth}
    \centering
    \includegraphics[width=\textwidth, trim={3cm, 9cm, 4cm, 10cm}, clip]{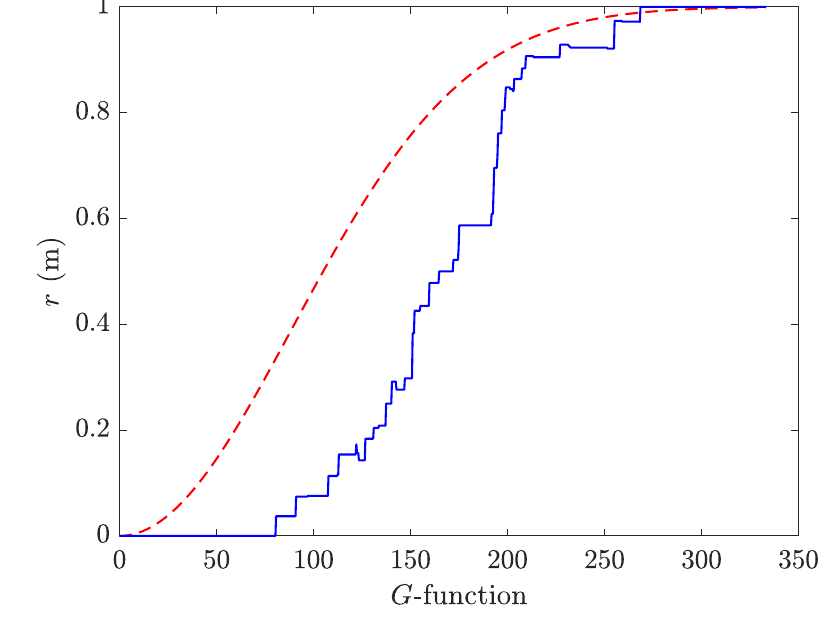}
    \label{fig:taxonomySmallRepulsion_G.pdf}
\end{minipage}%
\begin{minipage}{.2\textwidth}
    \centering
    \includegraphics[width=\textwidth, trim={3cm, 9cm, 4cm, 10cm}, clip]{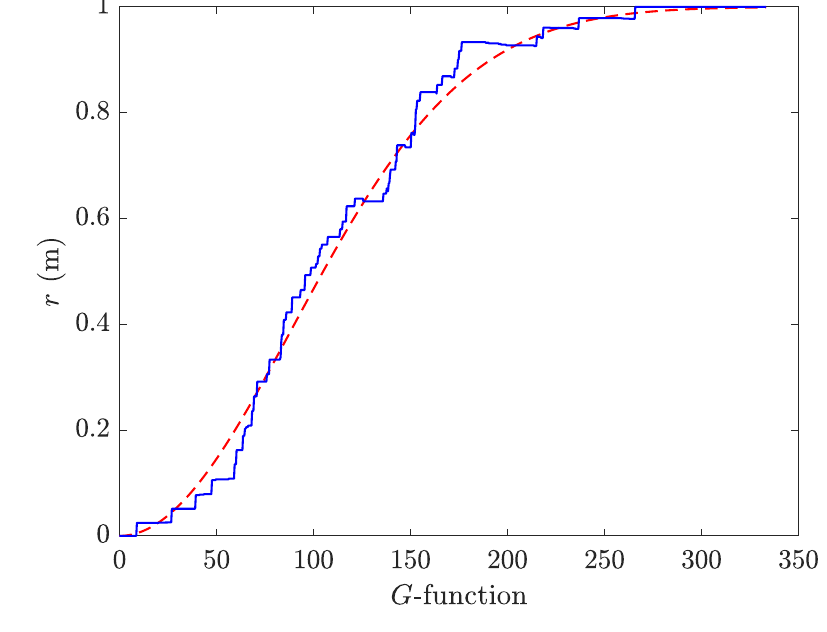}
    \label{fig:taxonomyPPP_G.pdf}
    \centering
\end{minipage}%
\begin{minipage}{.2\textwidth}
    \centering
    \includegraphics[width=\textwidth, trim={3cm, 9cm, 4cm, 10cm}, clip]{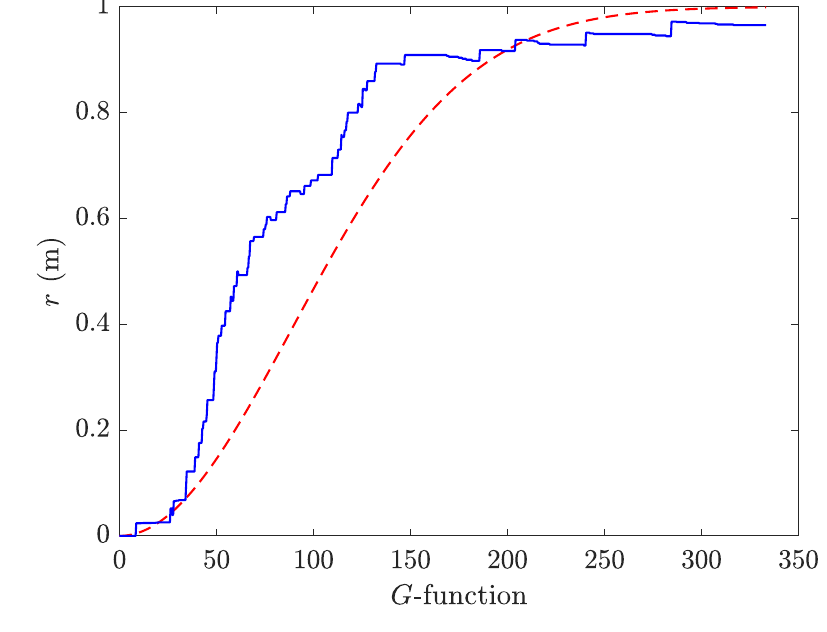}
    \label{fig:taxonomySmallAttraction_G}
\end{minipage}%
\begin{minipage}{.2\textwidth}
    \centering
    \includegraphics[width=\textwidth, trim={3cm, 9cm, 4cm, 10cm}, clip]{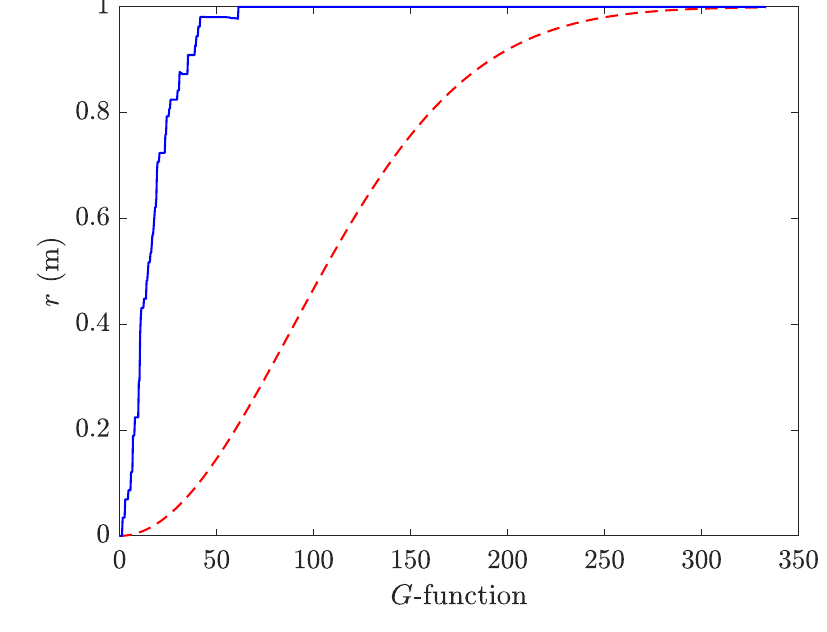}
    \label{fig:taxonomyHighAttraction_G}
\end{minipage}
\caption{\label{fig:PP_trichotomy_G}$G$-functions of the patterns of Figure \ref{fig:PP_trichotomy}. Extreme left: Perfectly regular network. Center: H-PPP. Extreme right: Highly clustered pattern.}
\end{figure*}

\begin{figure*}[p]
    \centering
\begin{minipage}{.2\textwidth}
    \centering
    \includegraphics[width=\textwidth, trim={3cm, 9cm, 4cm, 10cm}, clip]{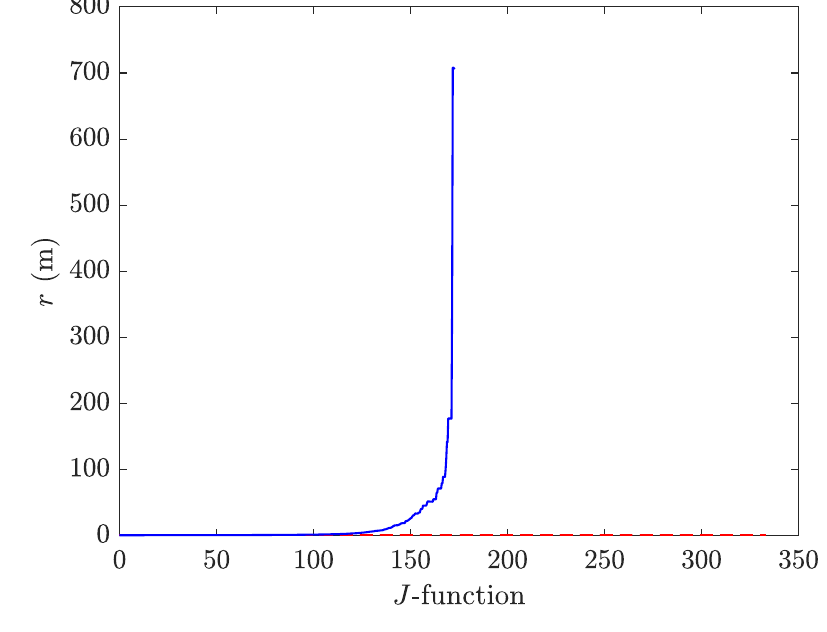}
    \label{fig:taxonomyregularNetwork_J.pdf}
\end{minipage}%
\begin{minipage}{.2\textwidth}
    \centering
    \includegraphics[width=\textwidth, trim={3cm, 9cm, 4cm, 10cm}, clip]{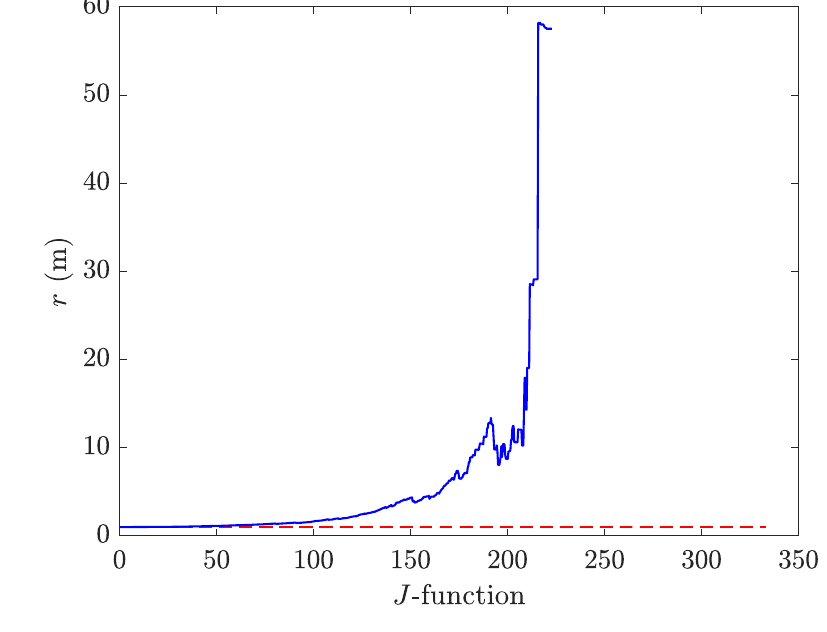}
    \label{fig:taxonomySmallRepulsion_J.pdf}
\end{minipage}%
\begin{minipage}{.2\textwidth}
    \centering
    \includegraphics[width=\textwidth, trim={3cm, 9cm, 4cm, 10cm}, clip]{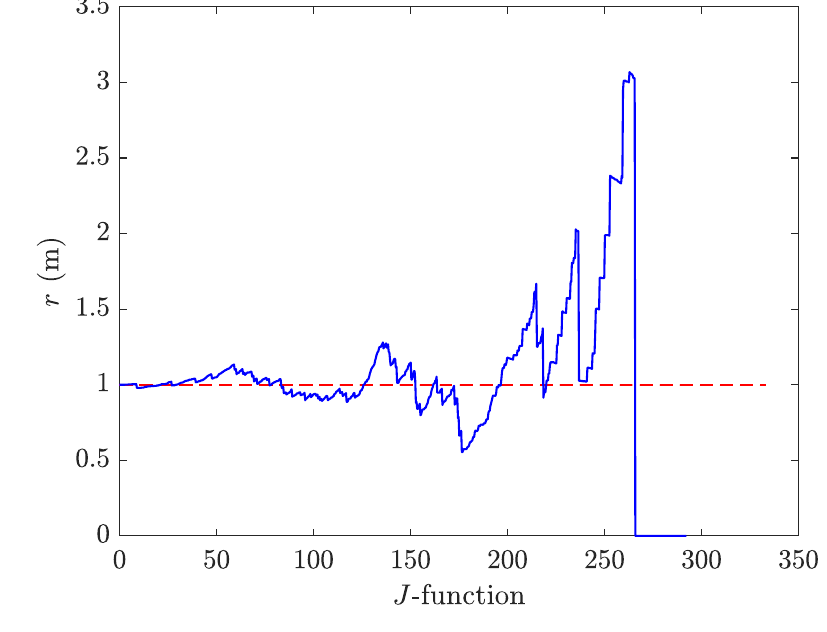}
    \label{fig:taxonomyPPP_J.pdf}
    \centering
\end{minipage}%
\begin{minipage}{.2\textwidth}
    \centering
    \includegraphics[width=\textwidth, trim={3cm, 9cm, 4cm, 10cm}, clip]{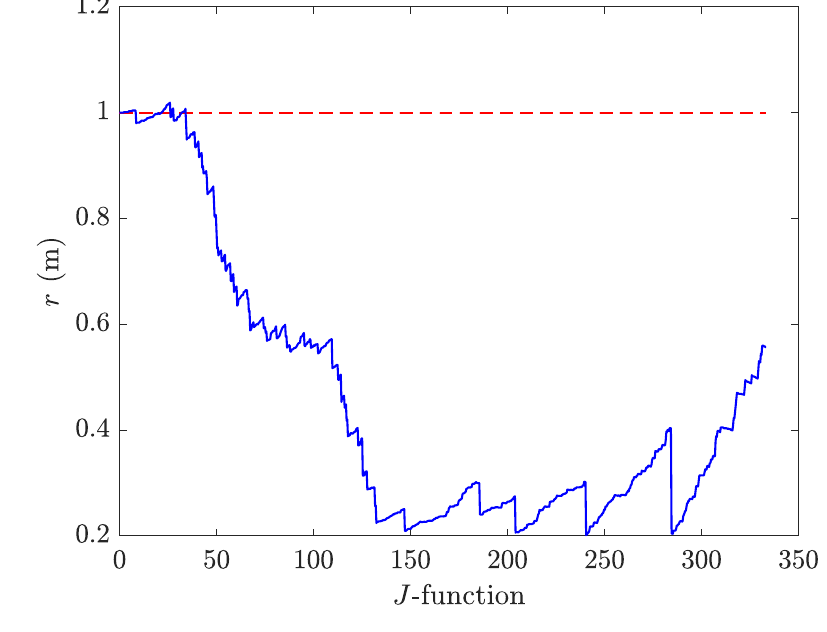}
    \label{fig:taxonomySmallAttraction_J}
\end{minipage}%
\begin{minipage}{.2\textwidth}
    \centering
    \includegraphics[width=\textwidth, trim={3cm, 9cm, 4cm, 10cm}, clip]{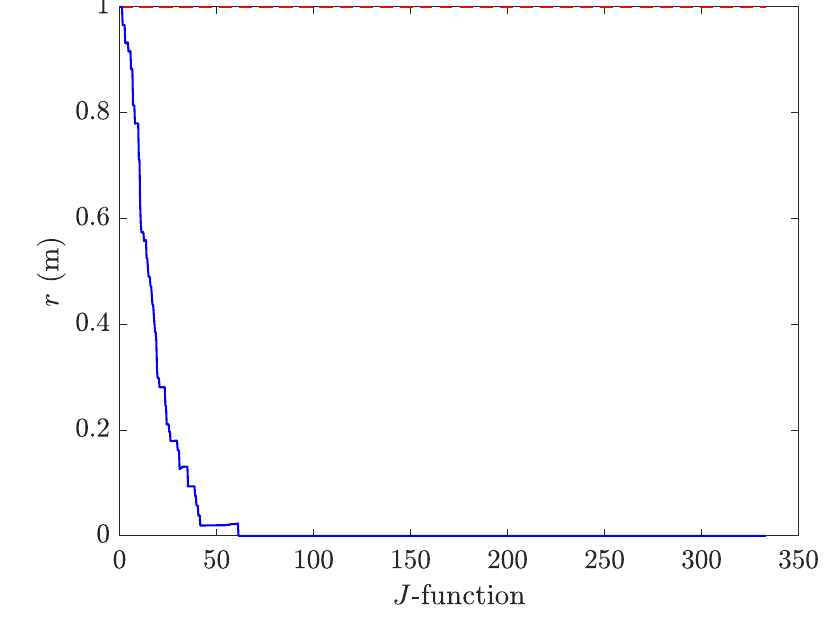}
    \label{fig:taxonomyHighAttraction_J}
\end{minipage}
\caption{\label{fig:PP_trichotomy_J}$J$-functions of the patterns of Figure \ref{fig:PP_trichotomy}. Extreme left: Perfectly regular network. Center: H-PPP. Extreme right: Highly clustered pattern.}
\end{figure*}

\subsection{Correlation}
Correlation quantifies the stochastic dependence between points in a spatial pattern of points. A clustered point pattern (right panels of Figure \ref{fig:PP_trichotomy}) has positive covariance, a completely random pattern has zero covariance (middle panel of Figure \ref{fig:PP_trichotomy}) and a regular pattern has negative covariance (left panels of Figure \ref{fig:PP_trichotomy}). Covariance is a second moment quantity. It is intrinsically linked to the count of pairs of points. The absence of correlation does not necessarily indicate independence between the points.

Correlation can be assessed using sample statistics such as the Morisita index \cite{HayesJames2017ANAf} or the index of dispersion \cite{P.GREIG-SMITH1952TUoR}, under the assumption of homogeneity \cite{R_spatstat}. Correlation is often studied in telecommunication networks through pairwise distance distributions. These metrics focus on the number of $r$-neighbors of a data point $x \in \Psi$, $n\left(d\left(x, \Psi \char`\\x\right)\right)$, i.e. the number points distant at most $r$ units away from $x$. The most used pairwise distance distribution is the Ripley $K$-function $K(r)$ \cite{KRipley}. It gives the expected number of $r$-neighbors of a data point randomly taken divided by the density $\lambda$. If the PP is stationary, its mathematical formulation is
\begin{align}\label{eq:K}
\begin{split}
    K(r) 
    &= \dfrac{1}{\lambda} \, \mathds E\left[n\left(d\left(u, \Psi \char`\\u\right) \leq r  \, | \, u \in \Psi\right) \right].\\
    \end{split}
\end{align}
$\lambda \, K$ is called non-regularized $K$-function. For a PPP, the points being independent, 
\begin{align}
   K_{pois}(r) &=  \dfrac{1}{\lambda} \, \mathds E\left[\sum_j \mathds 1\{||u-x_j|| \leq r\}\right] = \pi r^2.
\end{align}

The estimator $\widehat{K}(r)$ is
\begin{align}\label{eq:empiricalK}
    \widehat{K}(r) = \dfrac{\nu_2(\mathcal{B})}{N-1} \, \dfrac{1}{N} \,\sum_{i=1}^N\sum_{\substack{j=1 \\ j\neq i}}^N \mathds 1\{d_{ij} \leq r\}\, e_{ij}(r).
\end{align}
It corresponds to the average number of $r$-neighbors of a data point, normalized by the density of its $N-1$ neighbors included in the window $\mathcal{B}$ and modified by an edge correction weight. In general, we have $\widehat{K}^{regular}(r) < \widehat{K}^{independent}(r) < \widehat{K}^{clustered}(r)$ for the vast majority of $r$, as can be seen in Figure \ref{fig:PP_trichotomy_K}. This can be understood by the fact that a clustered point pattern has more close neighbors than in an independent pattern, which in turn has more close neighbors than in a regular pattern. The normalization thus allows comparison between datasets.

\subsection{Spacing}

Measuring the spacing between points in a point pattern provides additional information about the dependence between points. Spacing can be evaluated through sample statistics like the Clark-Evans aggregation index \cite{ClarkEvans} or via the Hopkins-Skellam test \cite{HOLGATEP1965Torb}. More insight is given by the popular empty-space distance distribution $F(r)$ (also called contact distance distribution) and nearest-neighbor distance distribution $G(r)$ defined for $r \geq 0$ for the stationary PP $\Phi$ respectively as
\begin{align}\label{eq:F}
    F(r) = \mathds{P}\left\{d(u, \Psi) \leq r\right\} = \mathds E \left[\mathds 1\left\{d(u, \Psi) \leq r\right\}\right],
\end{align}
\begin{align}\label{eq:G}
\begin{split}
    G(r) 
    &= \mathds{P}\left\{d(u, \Psi \char`\\ u) \leq r \, |\, \Psi \text{ has a point at } u\right\}\\
    \end{split}
\end{align}
where $u$ is an arbitrary location. Being cumulative distributions functions, $F(r)$ and $G(r)$ are zero a $r=0$, monotonically non-decreasing functions with a maximal value of 1. For a H-PPP,
\begin{align}\label{eq:Fpois}
    F_{pois}(r) = G_{pois}(r) = 1 - \exp(-\lambda\pi r^2).
\end{align}
If $N$ is the number of points of a stationary PP $\Psi$ falling in $\mathcal{B}$, estimators of $F(r)$ and $G(r)$ for a stationary PP are
\begin{align}\label{eq:empiricalF}
    \widehat{F}(r) = \dfrac{1}{M} \, \sum_{j=1}^M \mathds{1}\left\{d(u_j, \Psi) \leq r\right\}\,e_j
\end{align}
which is unbiased and
\begin{align}\label{eq:empiricalG}
    \widehat{G}(r) = \dfrac{1}{N} \, \sum_{x_i \in \mathcal{B}} \mathds 1 \left\{d(x_i, \Psi \char`\\ x_i) \leq r \right\}\,e_i
\end{align}
which is approximately unbiased. The number of random points $M$ can be chosen as high as wanted. $M > 10\,N$ is generally a good choice. $e_j$ and $e_i$ are the edge correction weights. As with the $K$-function, information about the data points can be obtained from the plot of the empirical $F$ and $G$-function (respectively in Figures \ref{fig:PP_trichotomy_F} and \ref{fig:PP_trichotomy_G} for the patterns in Figure \ref{fig:PP_trichotomy}), by comparing it to $F_{pois}(r)$ and $G_{pois}(r)$. Notice that because of its definition, which involves only data points, the empirical $G$-function is less smooth than the empirical $F$-function. For most values of $r$, we get: $\widehat{F}^{regular}(r) > \widehat{F}^{independent}(r) > \widehat{F}^{clustered}(r)$ and $\widehat{G}^{regular}(r) < \widehat{G}^{independent}(r) < \widehat{G}^{clustered}(r)$. Let us take a clustered pattern and an arbitrary point $u$. We expect that the probability of having a data point located at a given distance from $u$ is lower than for a random process. Similarly, since the probability of a data point belonging to a cluster is high, its nearest neighbor is located with high probability at a smaller distance than we would see for a random process.

As the interpretation of their empirical graphs already suggests, except for a completely random model, the nearest-neighbor and empty-space distances tend to have opposite behaviors. Another way to analyze these distances is to combine them through the van Lieshout-Baddeley $J$-function \cite{Jfunction}
\begin{align}\label{eq:J_def}
    J(r) = \dfrac{1-G(r)}{1-F(r)},
\end{align}
which is of course unitary in the case of a H-PPP. As can be expected from the previous considerations, $\widehat{J}^{regular}(r) > J_{pois}(r) = 1 > \widehat{J}^{clustered}(r)$ (see Figure \ref{fig:PP_trichotomy_J}). A value $J(100) = 3$ means that the distance threshold of 100 units is exceeded 3 times more often by the nearest-neighbor distances than by the empty-space distances. Since the $F$ and $G$-functions experience the same edge effects, we can consider that the $J$-function is insensitive to these edge effects. Neglecting all moments greater than second order, we have
\begin{align}\label{eq:J_approx}
    J(r) \approx 1 - \lambda (K(r) - \pi r^2),
\end{align}
showing the importance of the relationship between spacing and correlation.

\subsection{Repulsive Processes}
A very complete taxonomy of PPs capable of modeling antennas in a wireless communication network has already been presented in \cite{DiRenzoBible2021}. Detailed descriptions of classes of PPs, applications and mathematical expressions of most of the PPs introduced below can be found in \cite{Illianch3, R_spatstat, cressie1993, haenggi_2012}. The objective here is to remind the existing classes of PPs and to focus on those PPs for which mathematical expressions for the $F$, $G$ and $K$ functions have been demonstrated.

At the scale of a big city, repulsive processes can model networks of macro cells. Network providers typically attempt to optimize the deployment of macro antennas, often associated with high power, to provide coverage by limiting the number of base stations to be used. This logically leads to repulsive behavior.

We can already differentiate two types of repulsive processes: hard-core and soft-core repulsive processes. For hard-core processes, two points can never be closer than a deterministic distance $\delta$. Soft-core processes can be seen as obtained by progressively increasing the repulsion between points. There is always a non-zero probability of finding a pair of points closer than any small distance. For hard-core processes, the removal of two points that are too close can follow different schemes, giving birth to Matérn hard-core PPs (MHPP) of types I and II \cite{alma991004946749704066}, simple sequential inhibition process (SSI) \cite{hardcorePP} or Poisson hard-core process (PHCP). These processes are however intractable and approximations must often be used to compute mathematical expressions \cite{HaenggiMatern, CDDMatern}.

It is not possible to talk about repulsive processes without mentioning the large family of Gibbs processes. Gibbs processes are usually defined through a multivariate probability density function. Realizations of these processes can be simulated using Markov Chain Monte Carlo methods. Gibbs processes are constructed from the explicit interaction between points. Finite PPs can be represented as Gibbs PPs under certain conditions, but we generally restrict Gibbs PPs to processes with inhibition between points. An example of Gibbs hard-core PP is the Poisson hard-core process (PHCP). Amongst Gibbs soft-core PPs are the Strauss PP (SPP) which is defined similarly to the PHCP but with a non-zero probability of having two points with a distance smaller than $\delta$ or the Geyer-saturation PP (GSPP) which is a generalization of the SPP to also capture aggregation behaviors \cite{softcoreGibbs, Taylor_2012}.

Gibbs PPs as such suffer from some problems although they are widely used for PP inference. Closed-form prove difficult to obtain, complex Markov Chain Monte Carlo methods must be used to have a realization of the process \cite{Lavancier_2014}... This observation has gradually led to the emergence, over the past decade, of the idea of building determinantal PPs (DPP), described in \cite{Lavancier_2014, li2014statistical}. What follows is a summary of information needed to infer DPPs. DPPs are convenient models for repulsive macro cell networks inference because of their many advantages: they are very tractable, they have proven their validity for real networks, edge effects can be handled simply, they can be easily simulated... Each DPP $\Psi$ defined on a Borel set $B \in \mathbb{C}^d$ is defined through its covariance matrix $C^2 : B \to \mathbb{C}$. The name "determinantal" comes from the definition of the $n$th order product density function $\rho^{(n)} : B^n \to \mathbb{R}^+$ involving the determinant of the kernel:
\begin{align}
    \rho^{(n)}\left(x_1, \ldots, x_n\right) = \text{det} \left(C\left(x_i, x_j\right)\right)_{1 \leq i, j \leq n},\\
    \rho^{(n)}(x_1, \ldots, x_n) = 0 \text{ if } x_i = x_j \text{ for } i \neq j.
\end{align}
where $\left(x_1, \ldots, x_n\right) \in B^n$. Any Borel function $h : B^n \to \mathbb{R}^+$ can then be estimated through
\begin{align}
    &\mathds E \left[\sum_{\substack{X_1, \ldots, X_n \in \Psi \\ X_i\neq X_j}} h\left(X_1, \ldots, X_n\right)\right] = \\
    &\int_B \cdots \int_B \rho^{(n)}(x_1, \ldots, x_n) \, h(x_1, \ldots, x_n) \, dx_1 \cdots dx_n    
\end{align}
The DPP is stationary if its $n$-th order product density is invariant under translation. This is guaranteed if the kernel has the form $C(x,y) = C^{(0)}(x-y), \forall x, y \in \mathbb{R}^2$. $C_0$ is referred to as the covariance function of the DPP. As a consequence, the intensity measure is constant over $\mathds{R}^2$. For the case of a stationary DPP, the kernel only depends on the distance between the node pair.

The empty-space and nearest-neighbor distribution functions are given in \cite{li2014statistical}. Let us introduce some commonly-used DPPs:
\paragraph{Cauchy determinantal point process (CDPP)}
The Cauchy process is a stationary DPP with covariance function \cite{li2014statistical, Lavancier_2014, decreusefond2013perfect}
\begin{equation}
    C^{(0)}(x) = \dfrac{\lambda}{\left(1+||x||^2/\alpha^2\right)^{\nu+d/2}}, \qquad x \in \mathds{R}^d
\end{equation}
where $d=2$, $\alpha$ is the scale parameter and $\nu$ the shape parameter. The existence of the Cauchy DPP is guaranteed if $\lambda \leq \dfrac{\nu}{\pi\, \alpha^{2}}$.

The $K$-function is given by \cite{Lavancier_2014}
\begin{equation}
    K(r) = \pi r^2 - \dfrac{\pi \alpha^2}{2\nu+1} \, \left(1-\left(\dfrac{1}{1+r^2/\alpha^2}\right)^{2\nu+1}\right).
\end{equation}

It tends to a PPP when $\alpha \to \infty$.
\paragraph{Gauss determinantal point process (GDPP)}
A stationary PP $\Psi$ is a Gaussian DPP if it has as covariance function \cite{li2014statistical, Lavancier_2014, decreusefond2013perfect}
\begin{equation}
    C^{(0)}(x) = \lambda \, \exp\left(-||x||^2/\alpha^2\right), \qquad x \in \mathbb{R}^2
\end{equation}
where $\alpha$ measures the repulsiveness. The existence of the Gauss DPP model is guaranteed if $\lambda \leq \dfrac{1}{\pi\, \alpha^{2}}$.

The $K$-function is given by \cite{Lavancier_2014}
\begin{equation}
    K(r) = \pi r^2 - \dfrac{\pi \alpha^2}{2} \, \left(1-\exp\left(\dfrac{-2r^2}{\alpha^2}\right)\right).
\end{equation}

This is a PPP for $\alpha \to \infty$.

\paragraph{Generalized Gamma DPP}
This DPP shows a good fit with BS deployments but is less tractable. It is defined based on its spectral density rather than on a kernel \cite{Baccelli_DPP2}.

\paragraph{$\beta$-Ginibre point process ($\beta$-GPP)}
A zero value of $\beta$ corresponds to a PPP and a value of 1 corresponds to a GPP. A $\beta$-GPP is obtained from a GPP using a thinning approach. Each point of the GPP is kept independently with a probability $\beta$. Then, a rescaling ratio $\sqrt{\beta}$ is applied in order to maintain the original intensity. 

$\Psi$ is a $\beta$-Ginibre PP if the kernel is \cite{PP_cities, Ginibre_theory}
\begin{equation}
    C_{\beta}(x, y) = \lambda \, \exp\left(-\dfrac{\lambda\pi}{2\beta} \left(|x|^2+|y|^2\right)\right) \, \exp \left(\dfrac{\lambda\pi}{\beta} x \overline{y}\right)
\end{equation}
$\forall x, y \in \mathbb{C}$.
Contrarily to previous DPPs' kernels, the $\beta$-GPP kernel is complex. However, summary statistics have a tractable form. The $K$-function is given by \cite{Ginibre_theory}
\begin{equation}\label{eq:K_BGPP}
    K(r) = \pi r^2 - \dfrac{\beta}{\lambda} \, \left(1-\exp\left({-\dfrac{\lambda \pi r^2}{\beta}}\right)\right).
\end{equation}

The $F$-function is given by \cite{Ginibre_theory}
\begin{equation}\label{eq:F_BGPP}
    F(r) = 1- \prod_{k=1}^\infty \left(1-\beta\, \Tilde{\gamma}\left(k, \frac{\lambda \, \pi}{\beta}\, r^2\right)\right)
\end{equation}
and the $G$-function is given by \cite{Ginibre_theory}
\begin{equation}\label{eq:G_BGPP}
    G(r) = 1- \prod_{k=2}^\infty \left(1-\beta\, \Tilde{\gamma}\left(k, \frac{\lambda \, \pi}{\beta}\, r^2\right)\right)
\end{equation}
where $\Tilde{\gamma}(a, x) = \frac{\int_0^x \mathrm{e}^{-u}\, u^{a-1}}{\Gamma(a)}$ is the normalized lower incomplete gamma function.

The $J$-function is
\begin{equation}\label{eq:J_BGPP}
    J(r) = \dfrac{1-G(r)}{1-F(r)} = \dfrac{1}{1-\beta + \beta \, \exp\left(-\dfrac{\lambda \pi}{\beta}\, r^2\right)}.
\end{equation}

Mathematical expressions and approximations of metrics applicable to telecommunications can be found in \cite{GinibreNaoto14}.



\subsection{Cluster Processes}
Cluster processes are widely used to model small cells deployments, often clustered around hot spots in urban environments for capacity enhancement \cite{3GPP_Ktiers_smallcells, SBSWang16} or in indoor environments for coverage, to model drone networks \cite{Hayajneh2018PerformanceAO} or to model user distributions with hot spots \cite{peoplecluster}. Since this paper focuses primarily on macro cell deployments, this section merely provides a brief description of the existing cluster processes. 

Realizations of cluster usually follow the same scheme: First, a PP $\Psi_1$ of parent points is generated. Then, each point of $\Psi_1$ has a certain number of offspring points $\Psi_2$. Then, the cluster PP $\Psi$ is usually formed by some or all of the points of $\Psi_2$, only. Depending on the type of spatial distributions of $\Psi_1$ and $\Psi_2$, the resulting cluster PP $\Psi$ has specific properties and is called by a specific name. 

If the parent spatial distribution is a Poisson distribution, the PP belongs to the general class of Poisson cluster processes (PCP). They are often used due to their tractability. If each parent point produces a random number of offspring points that are independently and identically distributed with some spatial density function around the parent point, the PCP belongs to the subclass of Neyman-Scott cluster PPs \cite{Cauchy_Ghorbani}. The parent PP is then a H-PPP. Neyman-Scott cluster PPs have the advantage of having a good tractability: mathematical expressions have been derived, for example for the $K$-function \cite{GeneralizationNeymanScott} as well as for the nearest-neighbor distribution \cite{cressie1993, Heinrich88}. For the Matérn cluster process (MCP), the offspring points are generated in an exclusion ball of given radius $\delta$ centered on each parent point. If the probability density of offspring is an isotropic Gaussian density centered on the parent points, the PP is called Thomas cluster process (TCP). The TCP has the advantage to be one of the most tractable Neyman-Scott PP: mathematical expressions have been derived for its summary statistics \cite{Afshang_TCP}. The Cauchy cluster process is another type of Neyman-Scott cluster PP. The probability density of the offspring is a bivariate Cauchy distribution \cite{R_spatstat}. Another class of PCPs is the Gauss-Poisson PP for which the number of offspring is between zero and two and follows a Poisson binomial distribution \cite{haenggi_2012, GPPP}.

Cox cluster PPs, doubly stochastic Poisson processes or modulated Poisson processes, are Poisson processes with a random intensity function. This function is random because it generally depends on unobservable external factors as well as observable covariates. The offspring points are generated according to a I-PPP whose density is a realization of the random intensity function. Parent points are unobservable. Unlike I-PPPs, Cox cluster processes are stationary. The log-Gaussian Cox process (LGCP) is a subclass of Cox PPs for which the logarithm of the intensity function is a real-valued Gaussian process \cite{MOLLERJESPER2007MSfS}. The $\alpha$-stable Cox PP is another Cox PP for which the random field follows an $\alpha$-stable distribution \cite{alphastable}. For the shot-noise Cox PP, the random field is obtained from a general PP and the offspring points are then generated by an I-PPP whose density depends on the points already generated\cite{brix_1999}. Special cases of shot-noise Cox PP gave rise to Poisson-Gamma processes and shot-noise G-Cox processes \cite{moller_torrisi_2005}. Neyman-Scott PPs can also be considered as a special case of shot-noise Cox PP. 

At last, Poisson hole processes (PHPs) \cite{haenggi_2012} also belong to both Cox cluster PPs and PCPs. If $\Psi_1$ and $\Psi_2$ are two independent PPPs, the PHP is formed by taking points of $\Psi_2$ that are not included in holes of radius $\delta$ centered on parent points in $\Psi_1$. PHP can also be considered as repulsive PPs because the formation of holes forces the clustering of points.

\section{Other Point Processes}
Other families or subfamilies of PPs exist but are less interesting for inferring real networks. Perfect lattices such as the hexagonal \cite{hexagonal2016} or the square lattice \cite{InterferenceLargeNetworkHaenggi} were formerly used in telecommunications but have been replaced by the PPP or repulsive networks mentioned in this paper. However, it can be shown that when computing physical quantities, such as the SINR distribution, the perfect lattice gives an upper bound while the PPP gives a lower bound. Other models, such as the perturbed lattice or the combination of a lattice and a PPP, have been modeled, but are merely tractable. Permanental PPs, for example, are the reciprocal of determinantal PP, hence attractive, but are more interesting for quantum mechanical models \cite{Permanental_mccullagh_moller_2006, permanentalPP}.

Another modeling approach has been developed in \cite{IDT} and consists in approximating a motion invariant PP by a superposition of two conditionally independent I-PPPs exhibiting spatial repulsion or clustering. Measures such as SINR have been computed, but the consideration of two PPPs makes the mathematical expressions much more complex.

\section{Edge corrections}
Empirical data are always of finite dimensions and spread within a zone called window. Although the tractability of some spatial PPs is sufficient to work with a finite network, the edges of the window must be taken into account to compute sample or summary statistics, in order to reduce the bias caused by quantities related to points near the edges. For example, in a homogeneous PP, a point near the edge of a rectangular window has two to four times less neighbors than a point lying at the center of the window. Using edge corrections enables to somehow retrieve the lost information near the edges and to consider the empirical network as if it was infinite. The best correction choice depends on the number of data points and on the area of the window.

\begin{figure}[ht!]
    \centering
    \includegraphics[width=.9\linewidth, trim={3cm, 9cm, 4cm, 10cm}, clip]{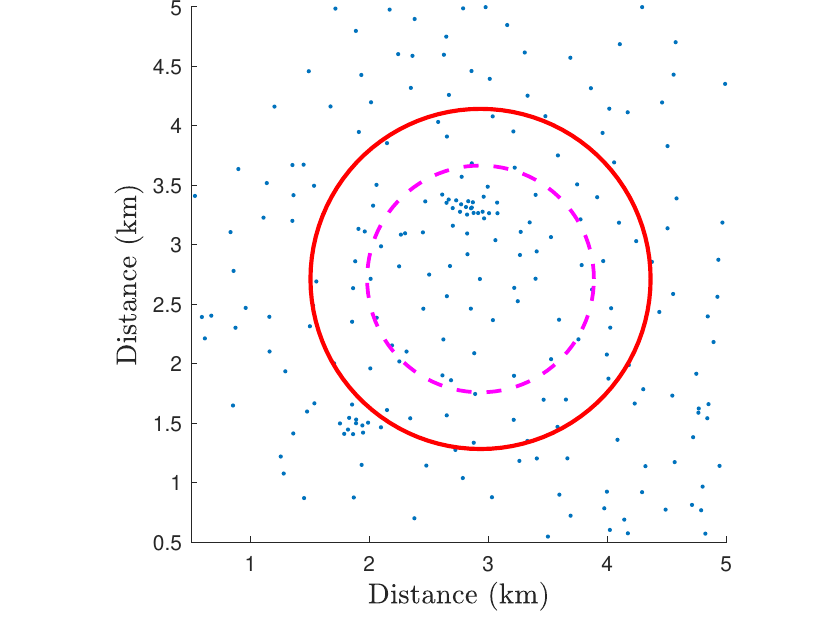}
    \caption{Illustration of the border correction. The solid red circle $\partial \mathcal{B}$ delimits the window $\mathcal{B}$ where all data points of interest are lying. The dashed purple circle corresponds to the sampling frame.}
    \label{fig:example_moving_border_correction_disc}
\end{figure}

The border correction \cite{R_spatstat} consists in restricting points to sampling frames smaller than the window size. For example, in Figure \ref{fig:example_moving_border_correction_disc}, the window considered $\mathcal{B}$ is the disk bounded by a red solid line $\partial\mathcal{B}$. We draw a sampling frame delimited by a purple dashed line. The distances $d_{ij}$ appearing in the estimator $\widehat{K}(r)$ \eqref{eq:empiricalK} are then considered only from data points $x_i$ located at least $r$ units away from the window boundary. The edge correction weight is thus
\begin{align}
    e_{ij} = \dfrac{\mathds 1\left\{d(x_i, \partial \mathcal{B} \geq r\right\}\, n}{\sum_{k=1}^n \mathds 1\left\{d(x_k, \partial \mathcal{B} \geq r\right\}}
\end{align}

Similarly for $\widehat{F}(r)$ \eqref{eq:empiricalF} and $\widehat{G}(r)$ \eqref{eq:empiricalG}, the edge correction weights are respectively
\begin{align}
    e_j = \dfrac{\mathds 1\left\{d(u_j, \partial \mathcal{B} \geq r\right\}\, m}{\sum_{k=1}^m \mathds 1\left\{d(u_k, \partial \mathcal{B} \geq r\right\}}
\end{align}
and
\begin{align}
    e_i = \dfrac{\mathds 1\left\{d(x_i, \partial \mathcal{B} \geq r\right\}\, n}{\sum_{k=1}^n \mathds 1\left\{d(x_k, \partial \mathcal{B} \geq r\right\}}.
\end{align}
The dashed purple circle of Figure \ref{fig:example_moving_border_correction_disc} becomes smaller and smaller as $r$ increases. The border correction is an easy method to implement, applicable to any window shape. It becomes more and more accurate as the size of the dataset increases. The only drawback is that part of the information is lost. We recommend to apply this correction when there are at least 80 data points in the window. Using the border correction does not further restrict the study because statistical inference for a network with less than 80 nodes becomes very imprecise. Other edge corrections can be applied, as the isotropic or Ripley's circumference method \cite{Yamada10, R_spatstat}, the toroidal method \cite{Yamada10}, the translation method \cite{R_spatstat}, area-based methods...

\section{Inference Method}
Based on the theory described in the previous section, we propose here a methodology to follow to infer actual network deployments and evaluate the goodness of fit. The methodology is applicable for any empirical stationary spatial distribution. It has been applied on macro cellular networks exhibiting repulsive behavior between nodes. 

\paragraph{Obtain antenna locations} The first step is to obtain the antenna locations. In some countries, this is made publicly available online. For example, the website \textit{Cartoradio} gives data about the French network via a direct download link. Sometimes the data is made available online, but no list of the entire network can be downloaded. This is for example the case in Belgium via the \textit{IBPT} website. In other countries like in the Netherlands or in Luxembourg, data is available online but no distinction is made depending on the technology, frequency or network provider, which is less suitable. A solution is then to download data from crowd-sourced services like \textit{Cell Mapper} or \textit{OpenCellid}, at the risk of missing data or incorrect locations. We strongly advise to use an adequate map projection if the coordinates are expressed with angular measurements, e.g. the geographic coordinate system (longitude and latitude). A widely used map projection for Western Europe is the Lambert~93 coordinate system.

\paragraph{Window selection and spatial stationarity} The choice of window is obviously important. Some edge corrections can only be applied for a rectangular or circular window (e.g., the Ripley circumference method or the toroidal method). Some irregular shapes or elongated windows may result in a large bias. If possible, it is advisable to take a compact surface and maximize the area-to-perimeter ratio. Regarding the window size, it is recommended to include a minimum of 80 antennas to have a good accuracy on the deduced model and to reduce the residual bias on the edge correction.

The present methodology being only applicable on stationary spatial distributions, a statistical test of stationarity has to be done.

\paragraph{Non-linear optimization of a given model} With models inferred based on summary statistics, the empirical functions $F$, $G$, and $K$ can first be calculated using the equations \eqref{eq:empiricalF}, \eqref{eq:empiricalG}, and \eqref{eq:empiricalK}, respectively. If there is clear repulsive or attractive behavior, some models can be discarded. This can be done by comparing the empirical summary functions to the H-PPP summary functions. By clear behavior, we mean an empirical summary function that is much higher or lower than the H-PPP summary function. The error distance $\Delta(\uptheta)$ between the summary statistics $S_\uptheta$ of the model with given model parameters $\uptheta$ and the empirical summary statistics $\widehat{S}$ can be calculated by the minimum contrast method
\begin{align}\label{eq:mcm_an}
   \Delta_S(\uptheta) = \int_{a}^{b} \left|\widehat{S}^p(r) - S^p_\uptheta(r)\right|^q\, dr,
\end{align}
where $a \leq r \leq b$ and $p, q > 0$ are exponents to choose. A squared Euclidean distance ($p = 1$ and $q = 2$) is generally selected.
Optimizing the model parameters then consists in minimizing the error $\Delta(\uptheta)$. In other words, to compute
\begin{align}
    \uptheta^* = \arg \min_\theta \Delta_S(\uptheta).
\end{align}
$\Delta_S(\uptheta)$ rarely has a closed-form expression. It is thus evaluated numerically via
\begin{align}\label{eq:mcm}
    \Delta_S(\uptheta) = \dfrac{1}{r_b-r_a}\,\sum_{i = a}^{b} \left|\widehat{S}^p(r_i) - S^p_\uptheta(r_i)\right|^q,
\end{align}
$a$ and $b$ being here indices.
The minimum contrast method is also called modified Cramér-von Mises method \cite{Cauchy_Ghorbani} when the integral in \eqref{eq:mcm} is computed over the whole positive real axis, that is, $r \leq 0$. In this paper, we propose to apply the non-linear optimization problem to the $F$-function for each candidate model.

\paragraph{Goodness-of-fit}
To evaluate the goodness-of-fit, i.e. the discrepancy between the fitted model and the empirical data, we propose to run a hypothesis test, often called 3SET for Summary Statistics Simulated Envelope Test. This test first consists in generating $M$ realizations of the model with model parameters $\theta^*$ and to compute the summary statistics $S_i$ for each realization $i$. Then, the upper envelope is $E_+(r) = \max\limits_{i=1,\ldots,M} S_i(r)$ and the lower envelope is $E_-(r) = \min\limits_{i=1,\ldots,M} S_i(r)$. These two pointwise envelopes form a gray region in which we will find $\widehat{S}$ if the PP correctly describes the empirical pattern. This test is statistically significant with a $p$-value of $2/(M+1)$. Another way to proceed is to take global envelopes, that is, to compute the maximum vertical deviation $D_{max}$ between the $M$ empirical $S$-function $S_{i|i=1\ldots M}$ and the theoretical $S$-function in order to obtain $E_+(r) = S_{\uptheta^*}(r) + D_{max}$ and $E_-(r) = S_{\uptheta^*}(r) - D_{max}$. In this case, the test is statistically significant with a $p$-value of $1/(M+1)$. We propose here to evaluate the goodness-of-fit for the $F$, $G$ and $K$ functions. The remaining candidate models will be those passing the 3SET test for these three summary statistics.

The best-fitting PP is then the one with the lowest value of $\Delta_F(\uptheta^*)$ among the remaining candidates. We then propose to evaluate the goodness-of-fit by also computing the distance between $G_\theta^*(r)$ and $\widehat{G}(r)$, between $J_\theta^*(r)$ and $\widehat{J}(r)$ and between $K_\theta^*(r)$ and $\widehat{K}(r)$. Other models use the $J$-function \cite{BGPP_Gomez}, the $K$-function \cite{Zhang_2021, li2014statistical} or the pair-correlation function \cite{Kibilda2016}. The problem with using the $J$-function is that it can behave erratically when the number of data points is too small. The $K$-function or the pcf are often used because they are simpler to compute than the spacing functions but it does not fully capture the dependence between points. The good fit with one summary statistics does not imply that all summary statistics will fit well. The idea here is to extend the goodness-of-fit to correlation and spacing functions.

\section{Experimental Results}

We have applied the described methodology to cellular networks in Belgium and in France. Several environments were considered:
\begin{itemize}
    \item Dense urban environment: the center of Paris, France, the center of Liège, Belgium
    \item Suburban environment: eastern suburban area of Paris, contained in the 93 and 94 departments.
    \item Rural environment: the Plateau de Millevaches, between Limoges and Clermont-Ferrand, in France and the north of Hainaut in Belgium
\end{itemize}
We have focused on one major operator in France and one major operator in Belgium for the GSM~900, UMTS~900 and LTE~1800 technologies. We have chosen a square window whose area enables to enclose at least 80 antennas. Data was downloaded in September 2021.

\begin{figure}[h!]
    \centering
    \includegraphics[width=.9\linewidth, trim={0cm, 0cm, 0cm, 0.8cm}, clip]{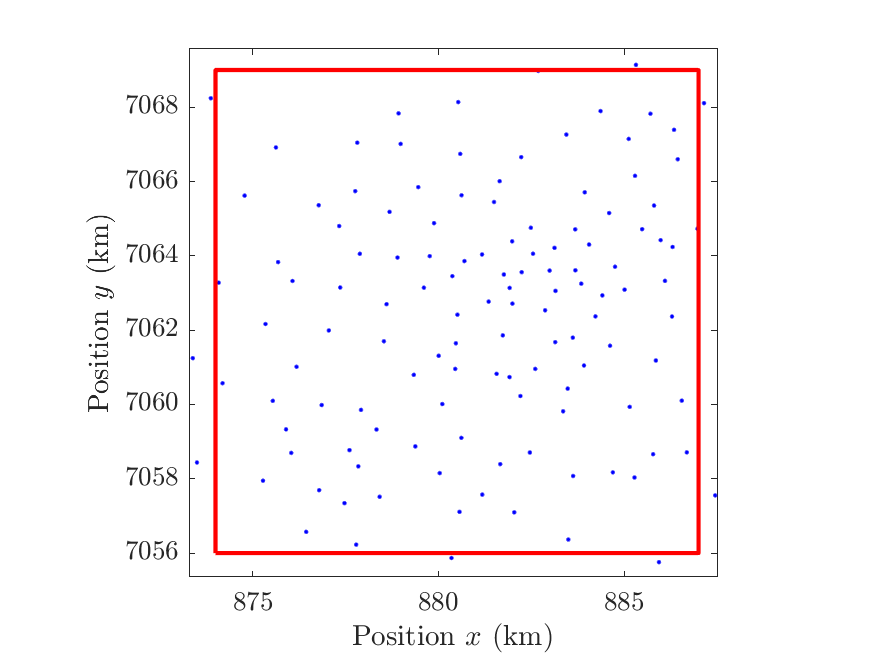}
    \caption{Locations of GSM~900 antennas of a major Belgian operator in the city of Liège. The axes are expressed in Lambert93 coordinates. The red square corresponds to the boundaries of the chosen window.}
    \label{fig:Belgium_PROXIMUS_GSM_900_x880.5_y7062.5_R6.5_Liege_zoom}
\end{figure}

We illustrate here the methodology for GSM~900 antennas in Liège, Belgium. The locations of the 119 enclosed antennas are shown in Figure \ref{fig:Belgium_PROXIMUS_GSM_900_x880.5_y7062.5_R6.5_Liege_zoom}. The estimated density is $\numprint[antennas/km^2]{0.70}$.

\begin{figure}[ht!]
    \centering
    \includegraphics[width=.9\linewidth, trim={3cm, 9cm, 4cm, 10cm}, clip]{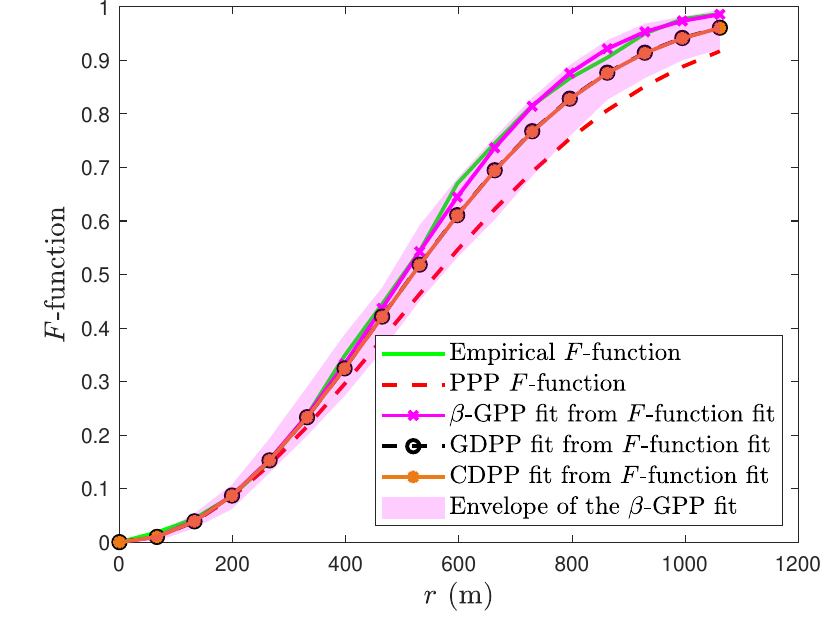}
    \caption{Empirical $F$-function and fitted $F$-functions for the $\beta$-GPP, GDPP and CDPP models.}
    \label{fig:Belgium_PROXIMUS_GSM_900_x880.5_y7062.5_R6.5_FfitfromF_envelope}
\end{figure}

\begin{figure}[ht!]
    \centering
    \includegraphics[width=.9\linewidth, trim={3cm, 9cm, 4cm, 10cm}, clip]{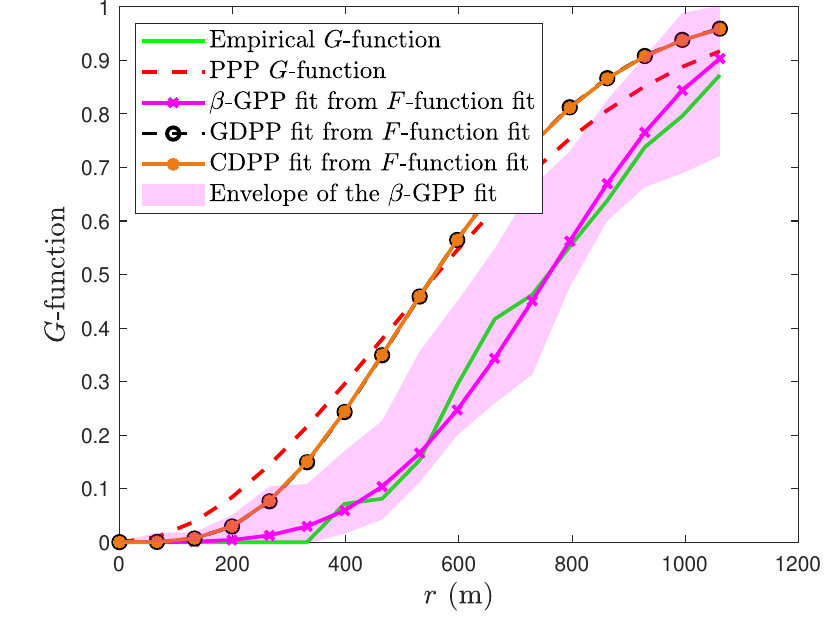}
    \caption{Empirical $G$-function and fitted $G$-functions for the $\beta$-GPP, GDPP and CDPP models.}
    \label{fig:Belgium_PROXIMUS_GSM_900_x880.5_y7062.5_R6.5_GfitfromF_envelope}
\end{figure}

When computing the empirical summary statistics, a clear repulsive behavior has been observed. We have therefore restricted the panel of candidate models to three repulsive and tractable PP: CDPP, GDPP and $\beta$-GPP. The minimum contrast method \eqref{eq:mcm} has been applied for the $F$-function, using $r_a = \numprint[m]{0}$, $r_b = \numprint[m]{1060}$, $p = 1$ and $q = 2$. $F$ and $G$ functions for the three models with the fitted model parameters are shown in Figures \ref{fig:Belgium_PROXIMUS_GSM_900_x880.5_y7062.5_R6.5_FfitfromF_envelope}, \ref{fig:Belgium_PROXIMUS_GSM_900_x880.5_y7062.5_R6.5_GfitfromF_envelope} and \ref{fig:Belgium_PROXIMUS_GSM_900_x880.5_y7062.5_R6.5_GfitfromF_envelope}, respectively. To evaluate the goodness-of-fit of the models, the pointwise envelopes where computed using $M = 39$ realizations. Only the  $\beta$-GPP model passed the test for the $F$, $G$, $J$ and $K$ functions with a $p$-value equal to 95\%. Its envelopes are also shown in Figures \ref{fig:Belgium_PROXIMUS_GSM_900_x880.5_y7062.5_R6.5_FfitfromF_envelope}, \ref{fig:Belgium_PROXIMUS_GSM_900_x880.5_y7062.5_R6.5_GfitfromF_envelope} and \ref{fig:Belgium_PROXIMUS_GSM_900_x880.5_y7062.5_R6.5_GfitfromF_envelope}. The fitted value of $\beta$ was found to be 0.91 with $\Delta_F(\beta = 0.91) = \numprint[]{9.36e-3}$, $\Delta_G(\beta = 0.91) = \numprint[]{2.94e-2}$, $\Delta_J(\beta = 0.91) = 1.03$ and $\Delta_K(\beta = 0.91) = \numprint[]{6.85e-2}$. 

\begin{figure}[ht!]
    \centering
    \includegraphics[width=.9\linewidth, trim={3cm, 9cm, 4cm, 10cm}, clip]{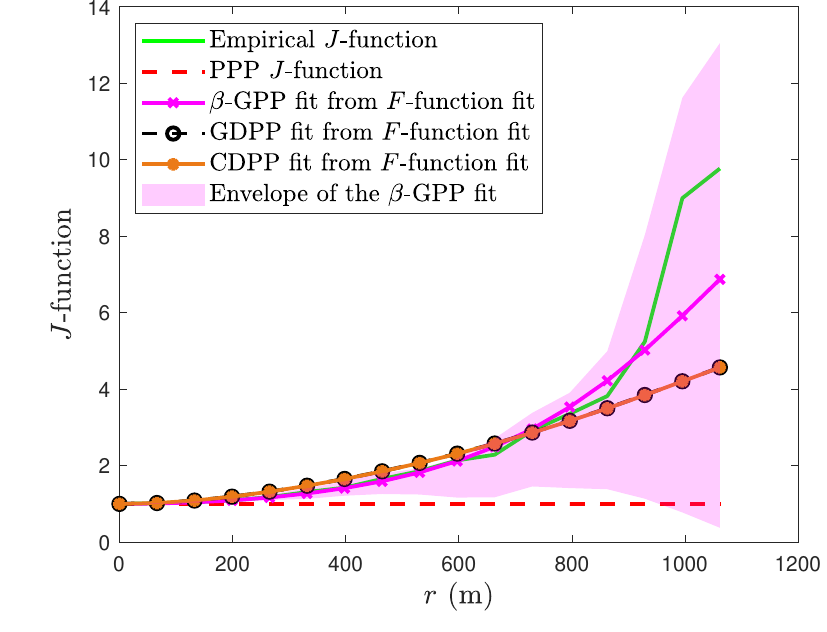}
    \caption{Empirical $J$-function and fitted $J$-functions for the $\beta$-GPP, GDPP and CDPP models.}
    \label{fig:Belgium_PROXIMUS_GSM_900_x880.5_y7062.5_R6.5_JfitfromF_envelope}
\end{figure}

For all the datasets considered in France and in Belgium, it appeared that the $\beta$-GPP model was the best adapted model each time. The $\beta$-values obtained for each dataset is summarized in Table \ref{tab:results_BGPP}. No clear correlation between the values of $\beta$ and the density, operator, technologies or place could be found with these datasets. However, the study of other technologies and operators suggests that there is a relationship between the beta value and the technology used, for a given type of environment.

\begin{table}[!ht]
\centering {
\caption{\label{tab:results_BGPP}{Fitted $\beta$ values of the $\beta$-GPP model for the considered datasets. "/" means that there is no or too few antennas of this type.}}
    \begin{tabular}{|l||c|c|c|}
        \hline
        Place & GSM~900 & UMTS~900 & LTE~1800\\
        \hline
        Liège & 0.91 & / & 0.86\\
        Hainaut & 0.15 & 0.18 & 0.13\\
        Paris & 0.95 & 0.54 & 0.31\\
        Eastern suburb of Paris & 0.50 & 0.67 & 0.66\\
        Millevaches & 0.17 & 0.83 & /\\
        \hline
    \end{tabular}}
\end{table}

\section{Conclusion}
Very often, SG studies are conducted with simple and well documented but sometimes too simplistic models, resulting in a description that is not very faithful to reality. However, spatial inference is a difficult exercise due to the lack of examples applied to real distributions, the need to resort to complex notions of SG, a lack of tractability of the models and a near absence of a "bestiary" detailing the main tractable PPs.

This paper tries to answer the expectations of a researcher wishing to model a real network by a PP. Under well-described starting assumptions about the network, we propose a methodology to follow. We review a range of tractable models that can be used as candidates, as well as the evaluation of goodness-of-fit.

This methodology is applied here to French and Belgian cellular networks. It is shown that among the models studied, the $\beta$-GPP model seems to describe the networks most accurately. No clear correlation between antenna density and beta value could be found.

\bibliographystyle{ieeetr}
\bibliography{bibli}

\end{document}